\documentclass{kapproc} 

\newread\epsffilein    
\newif\ifepsffileok    
\newif\ifepsfbbfound   
\newif\ifepsfverbose   
\newdimen\epsfxsize    
\newdimen\epsfysize    
\newdimen\epsftsize    
\newdimen\epsfrsize    
\newdimen\epsftmp      
\newdimen\pspoints     
\def\epsfbox#1{\global\def\epsfllx{72}\global\def\epsflly{72}%
   \global\def\epsfurx{540}\global\def\epsfury{720}%
   \def\lbracket{[}\def\testit{#1}\ifx\testit\lbracket
   \let\next=\epsfgetlitbb\else\let\next=\epsfnormal\fi\next{#1}}%
\def\epsfgetlitbb#1#2 #3 #4 #5]#6{\epsfgrab #2 #3 #4 #5 .\\%
   \epsfsetgraph{#6}}%
\def\epsfnormal#1{\epsfgetbb{#1}\epsfsetgraph{#1}}%
\def\epsfgetbb#1{%
\openin\epsffilein=#1
\ifeof\epsffilein\errmessage{I couldn't open #1, will ignore it}\else
   {\epsffileoktrue \chardef\other=12
    \def\do##1{\catcode`##1=\other}\dospecials \catcode`\ =10
    \loop
       \read\epsffilein to \epsffileline
       \ifeof\epsffilein\epsffileokfalse\else
          \expandafter\epsfaux\epsffileline:. \\%
       \fi
   \ifepsffileok\repeat
   \ifepsfbbfound\else
    \ifepsfverbose\message{No bounding box comment in #1; using
defaults}\fi\fi
   }\closein\epsffilein\fi}%
\def\epsfsetgraph#1{%
   \epsfrsize=\epsfury\pspoints
   \advance\epsfrsize by-\epsflly\pspoints
   \epsftsize=\epsfurx\pspoints
   \advance\epsftsize by-\epsfllx\pspoints
   \epsfxsize\epsfsize\epsftsize\epsfrsize
   \ifnum\epsfxsize=0 \ifnum\epsfysize=0
      \epsfxsize=\epsftsize \epsfysize=\epsfrsize
arithmetic!
     \else\epsftmp=\epsftsize \divide\epsftmp\epsfrsize
       \epsfxsize=\epsfysize \multiply\epsfxsize\epsftmp
       \multiply\epsftmp\epsfrsize \advance\epsftsize-\epsftmp
       \epsftmp=\epsfysize
       \loop \advance\epsftsize\epsftsize \divide\epsftmp 2
       \ifnum\epsftmp>0
          \ifnum\epsftsize<\epsfrsize\else
             \advance\epsftsize-\epsfrsize \advance\epsfxsize\epsftmp \fi
       \repeat
     \fi
   \else\epsftmp=\epsfrsize \divide\epsftmp\epsftsize
     \epsfysize=\epsfxsize \multiply\epsfysize\epsftmp
     \multiply\epsftmp\epsftsize \advance\epsfrsize-\epsftmp
     \epsftmp=\epsfxsize
     \loop \advance\epsfrsize\epsfrsize \divide\epsftmp 2
     \ifnum\epsftmp>0
        \ifnum\epsfrsize<\epsftsize\else
           \advance\epsfrsize-\epsftsize \advance\epsfysize\epsftmp \fi
     \repeat
   \fi
   \ifepsfverbose\message{#1: width=\the\epsfxsize,
height=\the\epsfysize}\fi
   \epsftmp=10\epsfxsize \divide\epsftmp\pspoints
   \vbox to\epsfysize{\vfil\hbox to\epsfxsize{%
      \includegraphics{#1}%
      \hfil}}%
\epsfxsize=0pt\epsfysize=0pt}%

{\catcode`\%=12
\global\let\epsfpercent=
\long\def\epsfaux#1#2:#3\\{\ifx#1\epsfpercent
   \def\testit{#2}\ifx\testit\epsfbblit
      \epsfgrab #3 . . . \\%
      \epsffileokfalse
      \global\epsfbbfoundtrue
   \fi\else\ifx#1\par\else\epsffileokfalse\fi\fi}%
\def\epsfgrab #1 #2 #3 #4 #5\\{%
   \global\def\epsfllx{#1}\ifx\epsfllx\empty
      \epsfgrab #2 #3 #4 #5 .\\\else
   \global\def\epsflly{#2}%
   \global\def\epsfurx{#3}\global\def\epsfury{#4}\fi}%
\def\epsfsize#1#2{\epsfxsize}
\def\eps@scaling{.95}
\def\epsscale#1{\gdef\eps@scaling{#1}}
\def\plotone#1{{\centering \leavevmode
    \epsfxsize=\eps@scaling\columnwidth \hfil \hbox{\epsfbox{#1}}\hfil}}

\let\footnote\savefootnote

\setcounter{tocdepth}{1}

\kluwerbib

\begin{document}



\articletitle[A Molecular Gas Survey of $z<0.2$ Infrared Excess, Optical
QSOs]{A Molecular Gas Survey of $z<0.2$ Infrared Excess, Optical
QSOs}


\author{A. S. Evans (SUNY, Stony Brook),\\ D. T. Frayer \& J. A. Surace (SIRTF
Science Center/Caltech), \\ and D.B. Sanders (Institute for Astronomy,
University of Hawaii)}



\begin{abstract}
Millimeter-wave (CO) observations of optically-selected QSOs are
potentially a powerful tool for studying the properties of both the
QSOs and their host galaxies.  We summarize here a recent molecular gas
survey of $z<0.17$ optical QSOs with infrared (IR) excess, $L_{\rm IR}
(8-1000\mu{\rm m}) / L_{\rm bbb} (0.1-1.0\mu{\rm m}) > 0.36$. Eight of
these QSOs have been detected to date in CO($1\to0$), and the derived
molecular gas masses are in the range $1.7-35\times10^9$ M$_\odot$. 
The high $L_{\rm IR}/L'_{\rm CO}$ of QSOs relative to
the bulk of the local ($z<0.2$) IR luminous galaxy merger population is
indicative of significant heating of dust ($L_{\rm IR}$) by the 
QSO nucleus and/or by massive stars created in the host galaxy with high
efficiency (i.e., per unit molecular gas, $L'_{\rm CO}$).
\end{abstract}


\section{Introduction}

Millimeter-wave (CO) observations of QSOs have not been a traditional
means of studying the underlying QSO host galaxies.  This is partially
due to the fact that millimeter-wave telescopes and arrays with large
($\geq$500 m$^2$) collecting areas and sensitive, low-noise receivers
have only been available within approximately the last decade; prior to
this time, single-dish and interferometric CO($1\to0$) observations of
``distant'' sources with faint mid- and far-IR flux densities, such
as the $z<0.16$ QSO Mrk 1014 (= PG 0157+001: Sanders et al. 1988a),
were time-consuming ventures. As a result, it is not surprising that
most of the advances in our understanding of QSOs have been achieved via
observations obtained at other wavelengths (primarily optical and radio).
The situation today is significantly different from what it was circa 1990
-- millimeter arrays operated by Caltech, the Berkeley-Illinois-Maryland
(BIMA) consortium, Nobeyama, and IRAM have five to six 10--15m diameter
dishes with state-of-the-art receivers, and upcoming facilities such as the
Smithsonian SubMillimeter Array (SMA), the Large Millimeter Telescope
(LMT), the Combined Array for Research in Millimeter Astronomy (CARMA),
and the Atacama Large Millimeter Array (ALMA) will provide additional
telescope availability and sensitivity with which to conduct routine,
large millimeter-wave surveys of distant galaxies in a variety of
different molecular tracers (e.g., CO, HCN).

For the past few years, we have made use of the Owens Valley
Millimeter Array (OVRO) to conduct a CO($1\to0$) survey of QSOs from the
Palomar-Green (PG) Bright Quasar Survey (Schmidt \& Green 1983). In the
context of this meeting, the motivation for undertaking such a survey
is three-fold: First, the rotational transitions of CO are tracers of
star-forming molecular gas, thus strong CO emission from QSOs is an
indication that their host galaxies have a significant cold molecular
component to their interstellar medium (ISM). Massive galaxies that
are known to contain such an ISM are spiral galaxies and ongoing
IR-selected mergers, which typically have molecular gas masses in
excess of $10^9$ M$_\odot$.  In contrast, optically-selected, massive
elliptical galaxies are intrinsically poor in molecular gas and dust,
with typical molecular gas masses $< 10^8$ M$_\odot$.  Second, in
addition to fueling star formation, molecular gas in the host galaxy is
a potential source of fuel for QSO activity. Thus, correlations between
the distribution, kinematics and amount of molecular gas and the level
of QSO activity may eventually aid in understanding the nature of
mass accretion processes of QSOs. Third, there exists the possibility
that some ultraluminous IR galaxy mergers (ULIGs: defined as having IR
luminosities, $L_{\rm IR} [8-1000\mu{\rm m}] \geq 10^{12}$ L$_\odot$)
are the evolutionary precursors of QSOs (Sanders et al.  1988b). In such
a scenario, ULIGs evolve from a molecular gas-rich, dust-enshrouded cool
(i.e., 25$\mu$m to 60$\mu$m flux density ratio, $f_{25}/f_{60} < 0.2$)
ultraluminous phase where vigorous star formation and accretion of
significant amounts of mass unto the supermassive nuclear black hole
have commenced, to a warm ($f_{25}/f_{60} \geq 0.2$) phase in which
AGN signatures are visible (i.e., bright nuclei with near-IR colors
consistent with a reddened QSO nucleus and Seyfert-like emission-line
spectrum), then finally to an UV-excess QSO.  This latter stage can only
occur after significant consumption or clearing of molecular gas and
dust from the nuclear region has occured, thus revealing the optical
QSO nucleus.  Given this, CO($1\to0$) observations of QSOs designed
to search for residual amounts of molecular gas from an earlier ULIG
phase may enable a direct comparison with the molecular gas content of
IR luminous galaxy mergers.

\section{IR-Excess, Optical QSOs}

In order to better facilitate comparisons with IR luminous galaxies, it
was necessary to select a QSO sample for which accurate determinations of
$L_{\rm IR}$ could be made (i.e., QSOs that have {\it IRAS}\footnote{I.e.,
the Infrared Astronomical Satellite} detections at mid and far-IR
wavelengths).  The sample was thus chosen from a $z<0.17$ IR-excess
(i.e. IR to ``big blue bump'' luminosity ratio, $L_{\rm IR}/L_{\rm bbb}
[0.10-1\mu {\rm m}] > 0.36$) sample of 18 PG QSOs compiled by Surace \&
Sanders (2001).\footnote{The sample was selected by Surace and Sanders for
the purpose of optical and near-IR imaging and is summarized in the Surace
\& Surace contribution to this proceedings.} The IR-excess criteria of
the sample thus selects the most likely ``transition'' candidates between
the dust-enshrouded ULIG phase and the optical, UV-excess QSO phase.

\begin{figure}
\includegraphics{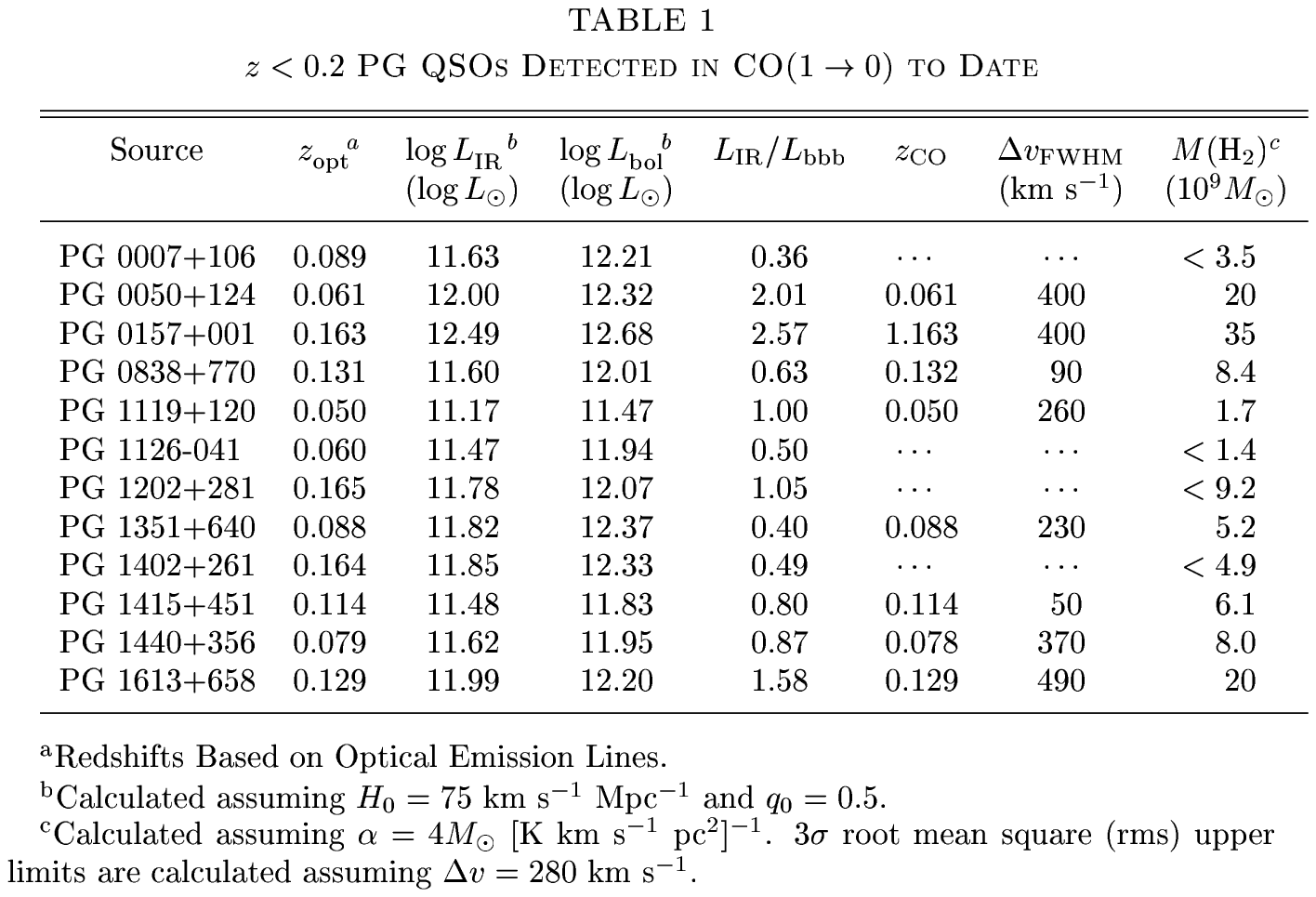}
\vspace{7.0cm}
\end{figure}

While selecting QSOs with IR excesses introduces a bias, it must be
remembered that 20--40\% of the bolometric luminosity, $L_{\rm bol}$,
of PG QSOs is emitted at IR wavelengths (Sanders et al. 1989), thus the
present sample of 18 QSOs simply populate one extreme of a fairly narrow
``IR luminosity fraction'' distribution.\footnote{See the M\"{u}ller
contribution to this proceedings for an updated discussion of PG QSO
spectral energy distributions.} Still, the {\it IRAS} detection rate of
spiral galaxies and ongoing mergers versus optically-selected elliptical
galaxies is very high, and if the host galaxies of QSOs are a mixture
of the above galaxy types, the ``IR-excess'' criteria will be biased
towards selecting QSOs hosts that are spiral galaxies and ongoing mergers.

\begin{figure}
\includegraphics{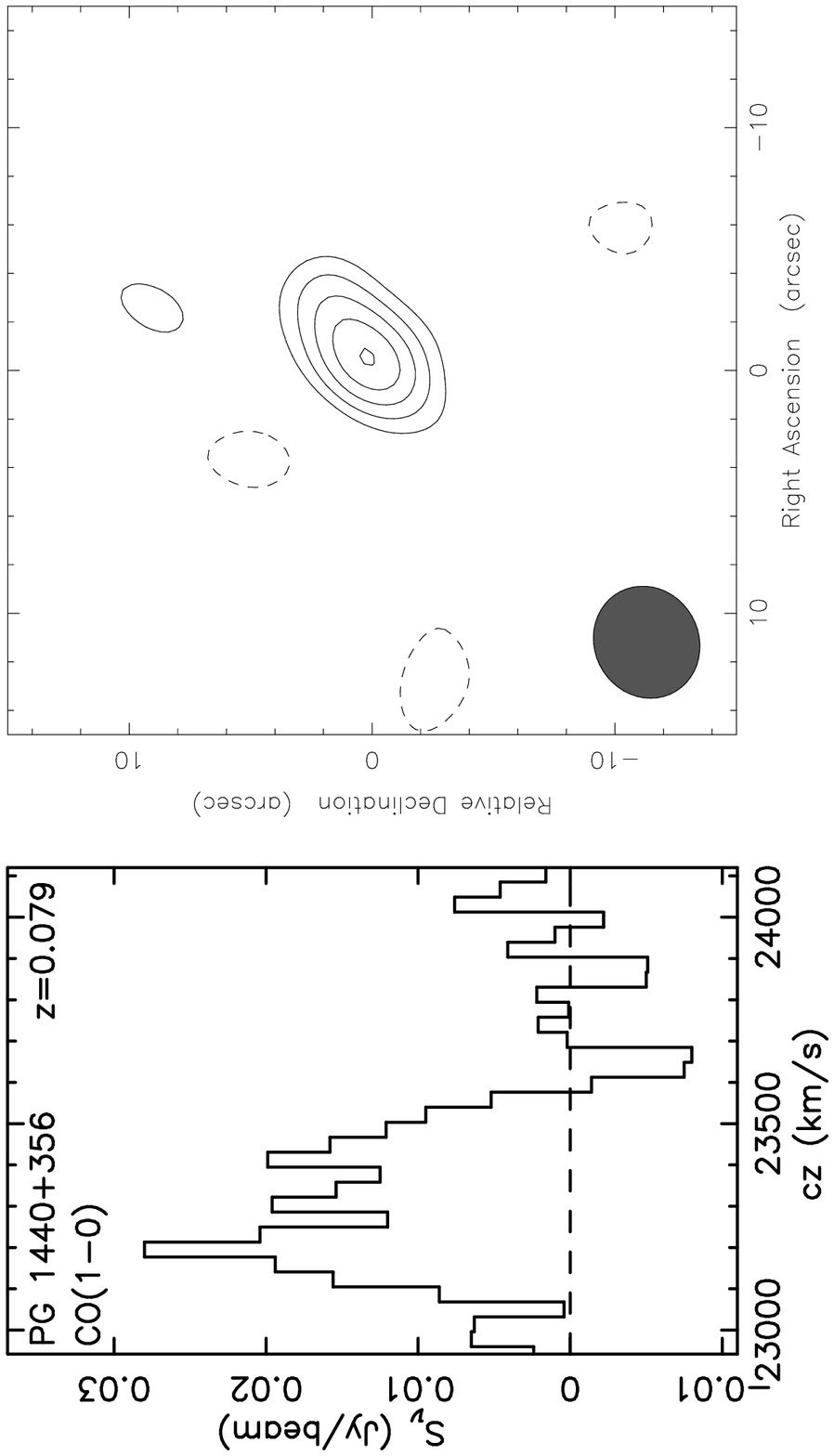}
\vspace{7.4cm}
\caption{CO($1\to0$) spectrum and integrated intensity map of the
IR-excess QSO PG 1440+356. For the map, contours are plotted as
1$\sigma \times (-2.3, 2.3, 3.3, 4.3, 5.3, 6.3)$; the peak intensity
is 0.016 Jy beam$^{-1}$, and corresponds to the position RA=14:42:07.48
dec=+35:26:22.33 (J2000.0).}
\end{figure}

The sample of 18 QSOs contains two QSOs (Mrk 1014 = PG 0157+001 and I
Zw 1 = PG 0500+124) that have been observed in CO multiple times with
different millimeter telescopes, and two QSOs (PG 0838+770 and 1613+658)
that have been observed once with the IRAM 30m telescope (e.g. Sanders et
al. 1988a; Barvainis et al. 1989; Alloin et al. 1992). Thus far in the
survey, PG 0838+770 and 1613+658 have been reobserved, and 8 additional
IR-excess QSOs have been searched for CO($1\to0$) emission for the
first time (see Table 1).  Two transits were typically done per source;
in terms of sensitivity, this corresponds to a 3$\sigma$ rms molecular
gas mass detection limit of $1\times10^9$ M$_\odot$ (assuming $\alpha =
4$ M$_\odot$ [K km s$^{-1}$ pc$^2$]$^{-1}$) for a $z \sim 0.1$ galaxy
with a CO velocity line width of 280 km s$^{-1}$.

\section{Molecular Gas Properties}

Table 1 lists several properties of the 12 IR-excess QSOs observed in CO
to date, and a CO($1\to0$) spectrum and integrated intensity map of one
QSO (PG 1440+356) is shown in Figure 1. Eight QSOs have been detected
thus far, with molecular gas masses in the range $1.7-35\times10^9$
M$_\odot$ and velocity line widths ranging from approximately 50 to 500
km s$^{-1}$. The molecular gas mass range of the detected QSOs indicates
the presence of host galaxies with massive cold molecular components,
but the upper limits of the remaining 4 QSOs yield inconclusive
results.

\begin{figure}
\includegraphics{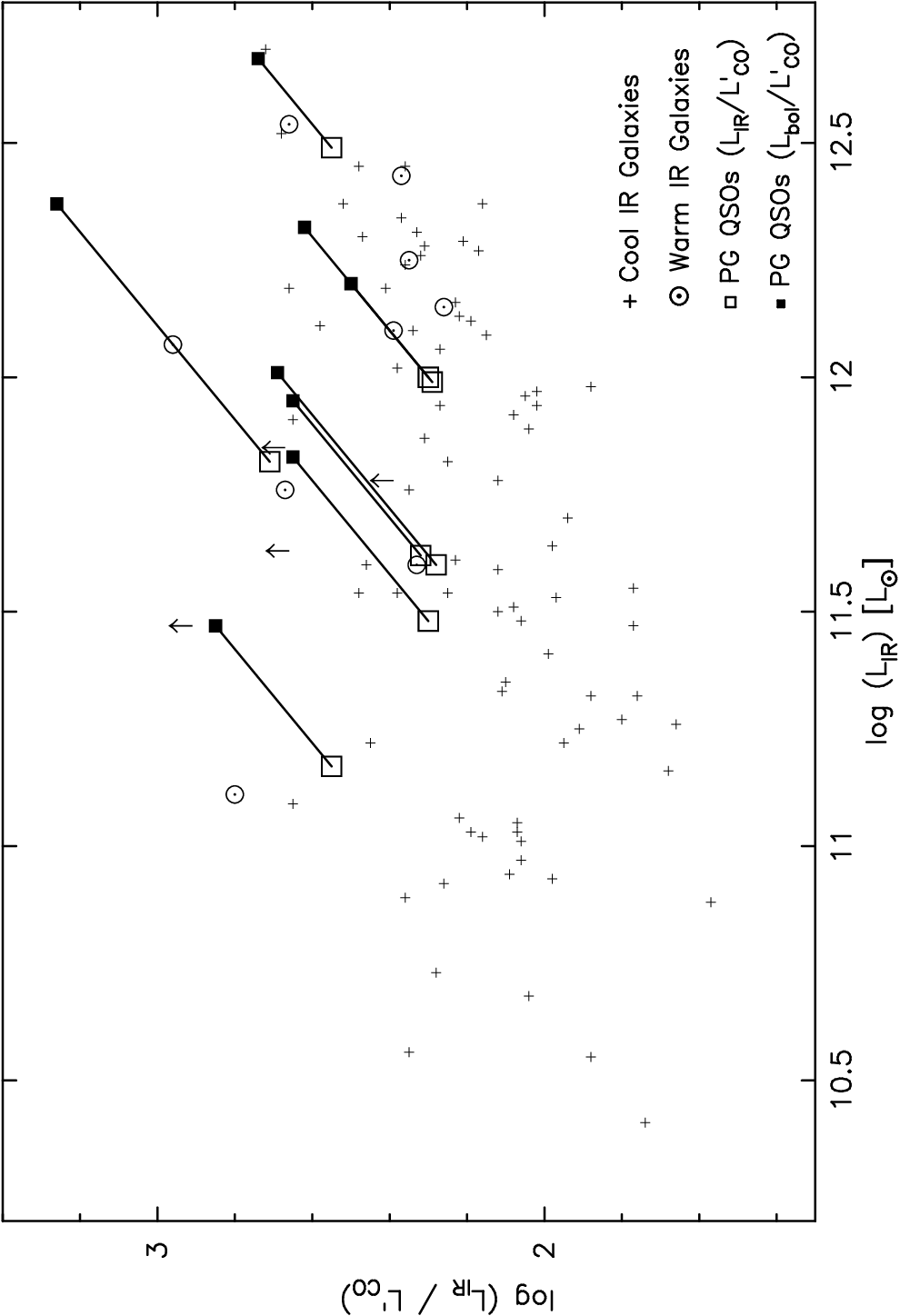}
\vspace{7.8cm}
\caption{A plot of $L_{\rm IR}/L'_{\rm CO}$ vs.
$L_{\rm IR}$ for the low-$z$ QSO sample, a flux-limited sample ($f_{60
\mu{\rm m}} > 5.24$ Jy) of IR luminous galaxies and a sample of
ultraluminous IR galaxies.  Arrows denote 3$\sigma$ lower limits on
$L_{\rm IR}/L'_{\rm CO}$ of PG 1126-041, PG 1202+281, and PG 1402+261.}
\end{figure}

In order to compare the CO($1\to0$) and IR properties of these QSOs and IR
luminous galaxies, $L_{\rm IR}/L'_{\rm CO}$ versus $L_{\rm IR}$ for both
a sample of cool and warm IR galaxies, as well as the IR-excess QSOs,
are plotted in Figure 2. The ratio $L_{\rm IR}/L'_{\rm CO}$ is commonly
referred to as the star formation efficiency; in starburst galaxies, it is
a measure of the cumulative luminosity of massive stars responsible for
heating the dust ($L_{\rm IR}$) relative to the amount of fuel available
for star formation ($L'_{\rm CO}$). The QSOs occupy the upper portion
of the $L_{\rm IR}/L'_{\rm CO}$ distribution of IR luminous galaxies
for a given value of $L_{\rm IR}$; this is an indication that the QSOs
contribute significantly to heating the dust in their host galaxies (thus
increasing $L_{\rm IR}$) and/or that the dust is heated by massive stars
formed in the host galaxy with high efficiency (per unit molecular gas
mass). The latter possibility may indicate that star formation accompanies
mass accretion unto supermassive nuclear black holes, and might support
recent observed correlations between black hole mass and both stellar bulge
mass and velocity dispersion in nearby quiescent galaxies (e.g. Magorrian
et al. 1998; Ferrarese \& Merritt 2000; Gebhardt et al. 2000).

Also plotted on Figure 2 are the positions the QSOs would occupy if
the optical QSO nucleus was completely enshrouded in dust (i.e., if
$L_{\rm IR} = L_{\rm bol}$, as is the case for ULIGs).  If QSOs are
the evolutionary products of ULIGs, and if ULIGs are powered primarily
by embedded QSO nuclei, then the area of Figure 2 covered by the lines
connecting $L_{\rm bol}/L'_{\rm CO}$ to $L_{\rm IR}/L'_{\rm CO}$ for
the QSOs show the possible paths that dust-enshrouded QSOs may follow
as they evolve towards UV-excess QSOs.

\section{The Future}

This CO survey is the first attempt to detect molecular gas in a complete
sample of QSOs, and it builds upon previous observations of $z<0.17$
observations of QSOs done by other groups. There are several ways in
which such a survey might be improved:

$\bullet$ The sample size is presently very small, and it is biased
towards IR-excess QSOs. A large CO survey of a volume-limited sample
of QSOs would give a more accurate assessment of the diversity in the
molecular gas content of QSOs.

$\bullet$ These observations were done in low-resolution mode ($4^{\prime
\prime}$ beam) for the simple purpose of making CO detections of
these QSOs.  Higher resolution ($0.5^{\prime \prime}$)  CO($2\to1$)
observations of the detected QSOs are required to determine the spatial
distribution and detailed kinematics of the molecular gas. Recent CO
observations by Schinnerer et al. (1998) have shown the CO in I Zw 1
(PG 0050+124) to be extended.



\begin{acknowledgments}

We are indebted to the OVRO staff and postdoctoral scholars
for their assistance. ASE
was supported by RF9736D and AST 0080881.

\end{acknowledgments}



%


\bibliographystyle{apalike}


\end{document}